\documentclass[review]{elsarticle}
\usepackage{lineno,hyperref}
\usepackage{longtable}
\modulolinenumbers[5]
\usepackage{multicol,color}
\usepackage{multirow}
\usepackage{multirow, array} 
\usepackage{booktabs}
\usepackage[ruled,vlined,linesnumbered,noresetcount]{algorithm2e}
\usepackage{setspace}
\newcolumntype{d}[1]{D{.}{.}{#1}}
\usepackage{float} 

\journal{Journal of \LaTeX\ Templates}

\biboptions{authoryear}

\biboptions{authoryear}

\usepackage{natbib}

\usepackage{rotating}

\usepackage[normalem]{ulem}
 \def\newblock{\ }%
\newcommand{\E}{{\cal E}}

\usepackage{amsmath,amsthm,amsopn,amstext,amscd,amsfonts,amssymb}
\usepackage{amsmath,amssymb}
\usepackage{amsthm}

 \newtheorem{definition}{Definition}
\newtheorem{exmp}{Example}

\begin{document}

\begin{frontmatter}


\title{The Concordance coefficient:  An alternative to the Kruskal-Wallis test} 
\author{Juan Francisco Monge}
\address{Operations Research Center, Universidad Miguel Hern\'{a}ndez. \\ Avda. de la Universidad, s/n, 03202-Elche (Alicante), Spain.\\
 E-mail addresses:   \, monge@umh.es\, (J.F. Monge)  }

\begin{abstract}
Kendall rank correlation coefficient  is used to measure the ordinal association between two measurements. In this paper, we introduce the Concordance coefficient as a generalization of the Kendall rank correlation, and illustrate its use to measure the ordinal association between quantity and  quality measures when two or more samples are considered. In this sense, the Concordance  coefficient  can be seen  as a generalization of the Kendall rank correlation coefficient and an alternative to the non-parametric mean rank-based methods  to compare two or more samples. A comparison of the proposed    Concordance coefficient and  the classical Kruskal-Wallis statistic is presented through a comparison of exact distributions of both statistics. 
\end{abstract}

\begin{keyword}
Kruskal- Wallis test - Kendall-$\tau$ correlation  - Kendall-$\tau$ distance - Linear Ordering Problem  
\end{keyword}

\end{frontmatter}



\section{Introduction}

 A parametric inference can be sometimes inappropriate. Assuming that the observations (samples) come from a certain distribution may not be very appropriate, since we may not have any type of information about the variable under study or evidence that implies a certain distribution in the observations.   Parametric statistics may also not be appropriate if the observations do not meet any of the basic assumptions for their use, for example; normality of data or having just a small number of observations.



Violation of the necessary assumptions in parametric statistics necessitates the use of non-parametric statistics. Non-parametric tests do not depend on the definition of a distribution function or statistical parameters such as mean, variance, etc. The use of non-parametric tests is also adequate when there are not enough observations available or when we are analyzing ordinal or nominal data.


Although the first steps in non-parametric statistics began earlier, it was in the 1930s when a systematic study in this field appeared.   \cite{Fisher1935} introduced the permutation test or randomization test as  a simple way to compute the sampling distribution for any test statistic under the null hypothesis that does not establish any effect on all possible outcomes. Over the next two decades some of the main non-parametric tests emerged, \cite{Friedman1940,Kendall1938_2,Kendall1938_1,Kruskal1958,Kruskal1952,Wilcoxon1947,Pitman1937,Wilcoxon1945}, among others.

The main advantages of the nonparametric test are: the data can be nonnumerical observations while they can be classified according to some criterion,   it is usually  easy to  calculate and does not make a hypothesis  about the distribution of the population from which the samples are taken.  We can also cite two drawbacks, the non-parametric test  is less precise than other statistical models and  it is  based  on the order of the elements in the sample and this order will likely stay the same even if the numerical data change.


There are many non-parametric tests in the literature, which can basically be classified into four categories depending on whether:   it is a test to   compare two, three o more related samples or a test for comparing related or unrelated samples.     Examples of the most used non-parametric tests in the literature for each of these four situations are the following:   the \emph{Wilcoxon signed-rank test}  (\cite{Wilcoxon1945})  for  comparing two related samples,  the \emph{Mann-Whitney (Wilcoxon) test} (\cite{Wilcoxon1947}) for  comparing two unrelated samples, the  \emph{Friedman test} (\cite{Friedman1940}) for comparing three or more related samples, the \emph{Kruskal-Wallis test} (\cite{Kruskal1952}) for comparing  three or more unrelated samples.

It is also possible to measure the degree of association of two variables through a non-parametric approach, in that sense we can mention  the Kendall rank correlation coefficient (\cite{Kendall1938_1}) and the Spearman rank correlation coefficient (\cite{Spearman1904}).

In (\cite{Aparicio2020}), the  authors introduce the \emph{Kendall-$\tau$ partition ranking};  given a ranking of elements of a set and given a disjoint partition of the same set, the \emph{Kendall-$\tau$ partition ranking} is  the induced  linear order of the subsets of the partition which follows from the given ranking of elements of a set. In this work we propose   to use the distance Kendall-$\tau$ as a concordance measure between the different samples in an  ordered  set  of observations. In this sense, the  measure we propose can be considered as an extension to the Kendall correlation coefficient when more than two samples are considered. The main difference between the proposed measure and the previous ones is the consideration of the Kendall distance instead   of ranks,  which use   classical methods. We also propose a significance test in order to determine when more than two samples come from the same distribution, and a comparison to the classical Kruskal-Wallis methods is presented.

The remainder of this paper is organized as follows. Section 2 presents the main features of the  Kendall correlation coefficient and the Kruskal-Wallis test. In  Section 3, we introduce   the Concordance coefficient while in  Section 4 the related statistical test is presented.  Section 5 presents some details of how the p-values have been calculated. Conclusions follow in Section 6. \ref{A:1} presents an example of the probability distribution of the Concordance coefficient, and \ref{A:2} deals with a  comparison between the distributions of the  Concordance coefficient and the Kruskal-Wallis statistic. Finally, critical values and exact p-values for the Concordance coefficient are presented in \ref{A:3}.

\section{Non-parametrical test}


This section presents the Kendall Rank Correlation Coefficient, a coefficient to measure the relationship between two samples ordinally, and the Kruskal-Wallis statistical test, which is a rank- based statistical test to measure whether different samples come from the same distribution, without assuming a given distribution for the population.

Only these two tests are presented in detail, since the test proposed in this paper uses the Kendall distance, 
and it  can be seen as an extension of the Kendall Rank Correlation Coefficient when more than two samples are considered and it is presented as an alternative to the Kruskal-Wallis statistical test.

\subsection{Kendall Rank Correlation Coefficient}

The Kendall rank correlation coefficient is a non-parametric measure of correlation. This measure is based on the Kendall-$\tau$ distance between two permutations of $n$ elements. The Kendall-$\tau$ distance ($d_{K-\tau}$) is defined as the number or pairwise disagreements between the two permutations $\pi_1$ and $\pi_2$. For instance, if we have three elements, the distance form permutation 123 to permutations 132, 231 and 321 is 1,2 and 3 respectively. The maximum number of disagreements that may occur between   two permutations of $n$ elements is $n(n-1)/2$, and in this case all the values of  permutation $\pi_1$  are in the reverse  order of $\pi_2$. 

The Kendall-$\tau$ rank correlation coefficient between permutations $\pi_1$ and $\pi_2$, denoted by $\tau$, is defined by 

$$ \tau = 1-  2 \frac{d_{K-\tau(\pi_1,\pi_2)}}{n(n-1)/2} .$$

The Kendall rank correlation  coefficient is used as a statistical test to determine whether there is a relationship or dependence  between two random variables, and as a non-parametric measure, it can be used with nominal or ordinal data. 

The main advantages of the Kendall Rank Correlation coefficient are: the data can be non-numerical observations if they can be ordered, it is easy to calculate, and, the associated statistical test does not assume a known distribution of the population from which the samples are taken.

\subsection{Kruskal Wallis Statistic}

The Kruskal Wallis test is a non-parametric statistical method to study whether different samples come from the same population. The test is the extension of the Mann-Whitney Test when we have more than two samples or groups. 
The following example illustrates the Kruskal Wallis test when comparing three samples.

\begin{exmp} 
. Let us assume that the effectiveness of three different treatments   ($A$, $B$, $C$)  has been measured for 6 individuals, two individuals assigned to each of the treatments, with the effectiveness of each treatment being measured ordinally. We could obtain the result shown in the following table, where, for example, the effectiveness of treatment A has been rated in first and third place. 
\begin{table}[h]\centering
\begin{tabular}{ccccccc}
 & $A$ & $B$ & $A$ & $C$ & $C$ & $B$ \\ \hline
 Rank & 1 & 2 & 3 & 4 & 5 & 6  \\ \hline 
 \end{tabular}
  \caption{Result for an experiment with 6 people and 3 treatments.}
  \label{tabla1:exmp}
\end{table}
\end{exmp} 

The Kruskal-Wallis statistic is determined by the difference between the ranks of the individuals in each category with the average range, so for example the average range of the test in this example is $\overline{R}=3.5$, while the average range of each of the three treatments are $\overline{R}_A=2$, $\overline{R}_B=4$ and $\overline{R}_C=4.5$. The Kruskal Wallis statistic (KW) s based on the calculation of the distance of each range to the average  range, which can be expressed as follows:


$$KW =  -3(n+1)+\frac{12}{n(n+1)}\sum_{i=1}^k \frac{R_i^2}{n_i} , $$

where  $n$ is the number of observations in  the $k$ samples, $n_i$ the number of observations in the $i$th sample and $R_i$ the sum of the ranks in the $i$th sample. In our example we can see the value of the statistic:
 
$$KW =  -3(n+1)+\frac{12}{n(n+1)}\sum_{i=1}^k \frac{R_i^2}{n_i} = -3(6+1)+\frac{12}{6(6+1)}\left (\frac{4^2}{2} + \frac{8^2}{2} + \frac{4.5^2}{2}  \right ) = 2. $$

Table \ref{t:prob_KW_222} shows the probability distribution of the Kruskal-Wallis  statistic for 3 treatments, each  with 2 patients. \ref{A:1} presents the Kruskal-Wallis statistic for all possible results in the experiment with 3 treatments and  2 people in each treatment.

\setlength{\tabcolsep}{7.3mm}
\renewcommand{\arraystretch}{.85}

\begin{table}[t]	\centering\footnotesize
\begin{tabular}{cc} 
   $KW$   &$ Prob$           \\ \toprule
4.57 & 0.06667 \\  \cmidrule(l){1-2}
3.71 & 0.13333 \\\cmidrule(l){1-2}
3.43 & 0.13333\\\cmidrule(l){1-2}
2.57 & 0.06667\\\cmidrule(l){1-2}
2.00     &  0.13333\\\cmidrule(l){1-2}
1.14 & 0.13333\\\cmidrule(l){1-2}
0.86 & 0.13333\\\cmidrule(l){1-2}
0.29 &0.13333\\\cmidrule(l){1-2}
0.00 & 0.06667 \\ \bottomrule[0.2mm]
\end{tabular}
\caption{Probability distribution for the Kruskal-Wallis $KW$ statistics. Sample size (2,2,2)}
\label{t:prob_KW_222}
\end{table}

\section{The  Concordance  coefficient $\tau$}

In (\cite{Aparicio2020}) the  authors introduce the \emph{Kendall-$\tau$ partition ranking};  given a ranking of elements of a set and given a disjoint partition of the same set, the \emph{Kendall-$\tau$ partition ranking} is  the induced  linear order of the subsets of the partition which follows from the given ranking of elements of a set.  The Kendall-$\tau$ partition ranking presents an ordinal alternative to the mean-based ranking that uses a pseudo-cardinal scale. For example, if we have a permutation  $\pi=(c|c|c|b|b|a|a|c|c)$ the elements in $V=A\cup B\cup C$, with $A=\{a,a\}, B=\{b,b\}$ and $C=\{c,c,c,c,c,\}$. The elements of  subsets $A$, $B$ and $C$  rank  6.5, 4.5 and 4.6 respectively. The   \emph{Kendall-$\tau$ partition ranking} is the problem to  find the closest permutation to $\pi$  which verifies  that all the elements belonging to the same  subset of the partition are consecutively listed, where the distances are  measured with the Kendall-$\tau$ distance  formula. In this example,  
 the Kendall-$\tau$ distance is 8 ($d_{K\text{-}\tau}=8$), the less number  of  pairwise disagreements form $\pi$ that allows  all the elements belonging to the same  subset of the partition to be consecutively listed, $\rho=(c|c|c|c|c|b|b|a|a)$. The Kendall-$\tau$ distance from a permutation $\pi$ is given 
	$$ d_{K\text{-}\tau} = \displaystyle \min \{ d_{K\text{-}\tau}(\rho, \pi):\text{ elements in }V_r\text{ are consecutively listed  in }\rho,\,\, \forall r \}.$$
	
 This distance is also called the  disorder of permutation $\pi$. For  the  calculation of the disorder of a permutation of elements, in  \citet{Aparicio2020}  the authors establish that the distance or disorder of a permutation of elements
$\pi = (a|a|b|b|a|c|a|b|c|\cdots|c|a|b)$   is given by the solution of the Linear Ordering Problem (LOP) with the preference matrix $M$, where the element $m_{ab}$  of matrix $M$ indicates the number of times that an element $a$ of sample $A$ precedes an element $b$ of sample $B$ in the order  $\pi$.  The solution of the linear ordering problem gives us a new order in the elements of  $\pi$, the closest to  $\pi$, in which all the elements belonging to the same sample are listed consecutively.
  The book \cite{Marti2011} provides a exhaustive study of the Linear Ordering Problem. 
 
  
 The authors  (\cite{Aparicio2020}) present the properties of the  \emph{Kendall-$\tau$ partition ranking} and compare it with a classical mean and median- based rank approaches. Those properties are extracted form social choice theory and adapted to a partition ranking, see (\cite{arrow1951,Kemeny1959,zahid2015}) .  Two of these properties are only  true for the   \emph{Kendall-$\tau$ partition ranking}: the \emph{Condorcet}  and  \emph{Deletion Independence} properties. The Condorcet property  establishes that  the most preferred subset must be listed before any other in any ranking; and the Deletion Independence property establishes that if any subset is removed, then the induced order of subsets does not change. In permutation $\pi=(c|c|c|b|b|a|a|c|c)$ the set $C$ is a condorcet winner, the most preferred set, but $B$ has a lesser mean rank value  than set $C$ if set $A$ is not considered in the comparison, therefore, the permutation $\pi=(c|c|c|b|b|a|a|c|c)$ gives an example where ranking  subsets from ranks is no very reliable. 

From \cite{Aparicio2020},  the maximum number of disagreements that may occur into a  permutation of $n$ elements (where the elements are classified in $k$ subsets  $V_1, V_2,\ldots V_k$ of sizes $n_1,n_2,\cdots,n_k$ respectively) is $\sum_{r<s} n_r \, n_s   -  (GP_{b}
+\sum_{r<s}\displaystyle\lfloor\frac{n_r n_s}{2}\displaystyle\rfloor )$, where $GP_{b}$ is the Generalized Pentagonal Number of $b$, and  $b$ the number of subsets $V_k$ with odd cardinality.  The Generalized Pentagonal number $GP_{b}$ is $$GP_b=\left\{ \begin{array}{ll}\displaystyle\frac{\ell(3\ell -1)}{2} & b=2\ell\\  & \\\displaystyle\frac{\ell(3\ell +1)}{2} & b=2\ell +1\\ \end{array} \right. .$$

This maximum number of  disagreements (the maximum disorder)   in a permutation $\pi$  of elements allows us  to define a relative disorder coefficient  of permutation $\pi$ as
	
	$$ disorder(\pi) = \frac{\displaystyle d_{K\text{-}\tau}(\pi)}{\displaystyle\sum_{r<s} n_r \, n_s   -  (GP_{b}
+\sum_{r<s}\displaystyle\lfloor\frac{n_r n_s}{2}\displaystyle\rfloor )} .$$

\begin{definition}
	We define the  Concordance coefficient ($\tau$) of permutation $\pi$ as
	$$\tau= 1-disorder(\pi) = 1- \frac{\displaystyle d_{K\text{-}\tau}(\pi)}{\displaystyle \sum_{r<s} n_r \, n_s   -  (GP_{b}
+\sum_{r<s}\displaystyle\lfloor\frac{n_r n_s}{2}\displaystyle\rfloor )} .$$
\end{definition}

 The   Concordance coefficient ($\tau$)   provides a measure of independence in the $k$ samples, where $\tau$ is a value between 0 and 1, taking the value of 1 when there is a total order between the samples and 0 when the disorder is maximum. In this sense, the Concordance coefficient $\tau$ can be seen as a generalization of the Kendall Rank Correlation Coefficient when we have more than two samples.
 

\setcounter{exmp}{0}
\begin{exmp}[Cont.] 
Continuing with the data in Example 1, the results of the experiment provide the following order or permutation of the treatments $\pi=(a|b|a|c|c|b|)$.
\end{exmp}

Given the order of individuals $\pi=(a|b|a|c|c|b|)$, the ordering between individuals that leaves individuals with the same treatment together is ordination ($a$ $a$ $b$ $b$  $c$ $c$ )  or the ordination ($a$ $a$ $c$ $c$  $b$ $b$ ). Both ordinations only need 3 pairwise disagreements from the permutation $\pi$.  In order to find the permutation of elements (equal elements listed consecutively) closer to a given permutation, it is sufficient to solve the Linear Ordering Problem (LOP) with the preference matrix defined above, in this example said matrix is:


\setlength{\tabcolsep}{1mm}
\begin{center}
\begin{tabular}{cccc}
&\,\,\, \,\, $A$ & $B$ & $C$\,\,\,\, \\ 
$A$&\multirow{3}{*}{  $ \left(\begin{array}{r}  $--$ \\1  \\  0 \end{array}\right.$}  &\multirow{3}{*}{  $ \begin{array}{c}3   \\ $--$  \\  2\end{array}$}   &\multirow{3}{*}{  $ \left.\begin{array}{c}4   \\2  \\  $--$ \end{array}\right)$}  \\
$B$&&&\\
$C$&&&\\
 \end{tabular},\end{center}

 where each element of the matrix $m_{ij}$ represents the number of times an individual of a treatment $i$ precedes an individual of the treatment $j$. The solution of the LOP is the permutation of treatments, which maximizes the preferences of order in the experiment, that is, in this example, the permutations of treatments $(A\ B\ C)$ or $(A\ C\ B)$ retain 9 preferences expressed in the order of individuals represented by the permutation  $\pi$.  Therefore, the permutation distance $\pi$ to a total order between treatments is $\sum_{i<j} n_i n_j -9 =3$.  At this distance, the number of pairwise disagreements needed in a permutation of elements to reach a permutation that establishes a total order between treatments, the authors of the work  \citep{Aparicio2020}   denominate disorder of permutation.\footnote{If the number of samples is small, we can evaluate all the possibilities in order to obtain the solution of the Linear Ordering Problem, for example, if we have 3 samples the number of feasible solutions for the LOP  is $3!=6$.}



Then, the relative  disorder of permutation $\pi$ can be evaluated as 

$$ disorder(\pi) = \frac{\displaystyle d_{K\text{-}\tau}(\pi)}{\displaystyle\sum_{r<s} n_r \, n_s   -  (GP_{b}
+\sum_{r<s}\displaystyle\lfloor\frac{n_r n_s}{2}\displaystyle\rfloor )} =\frac{\displaystyle 3}{\displaystyle 12  -  (0
+6  )} = \frac{3}{6}=\frac{1}{2},$$

and the Concordance coefficient\footnote{\ref{A:1} presents the disorder and the Concordance coefficient  for all possible results in the experiment with 3 treatments and  2 people in each treatment.}

$$\tau= 1-disorder(\pi) = 1-\frac{1}{2}=\frac{1}{2}.$$ Notice that no set of this example has odd cardinality, therefore the pentagonal number $GP_0=0$.

\begin{figure}[t]\centering
\begin{tabular}{cc}
		\includegraphics[width=4.95cm]{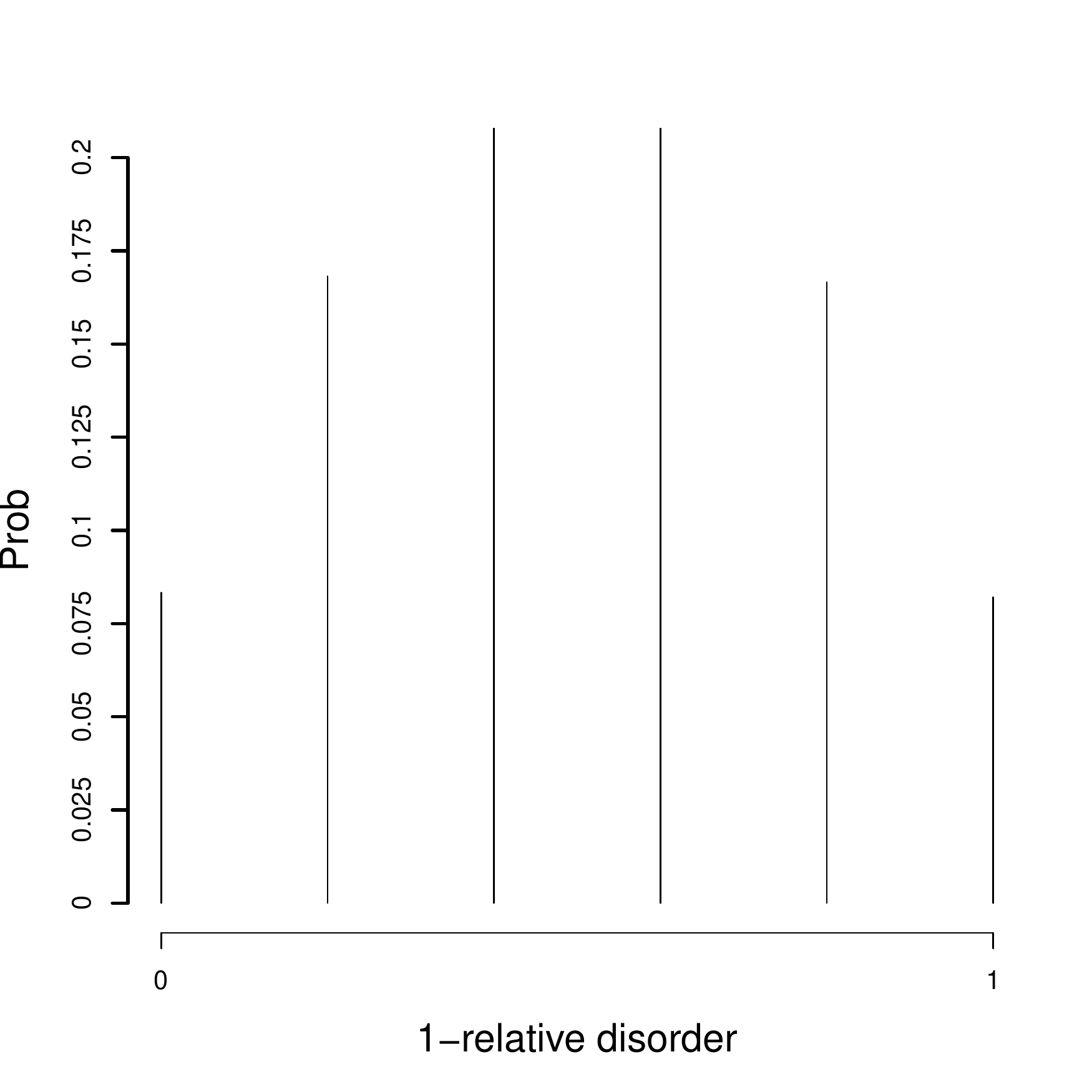}  &\includegraphics[width=4.95cm]{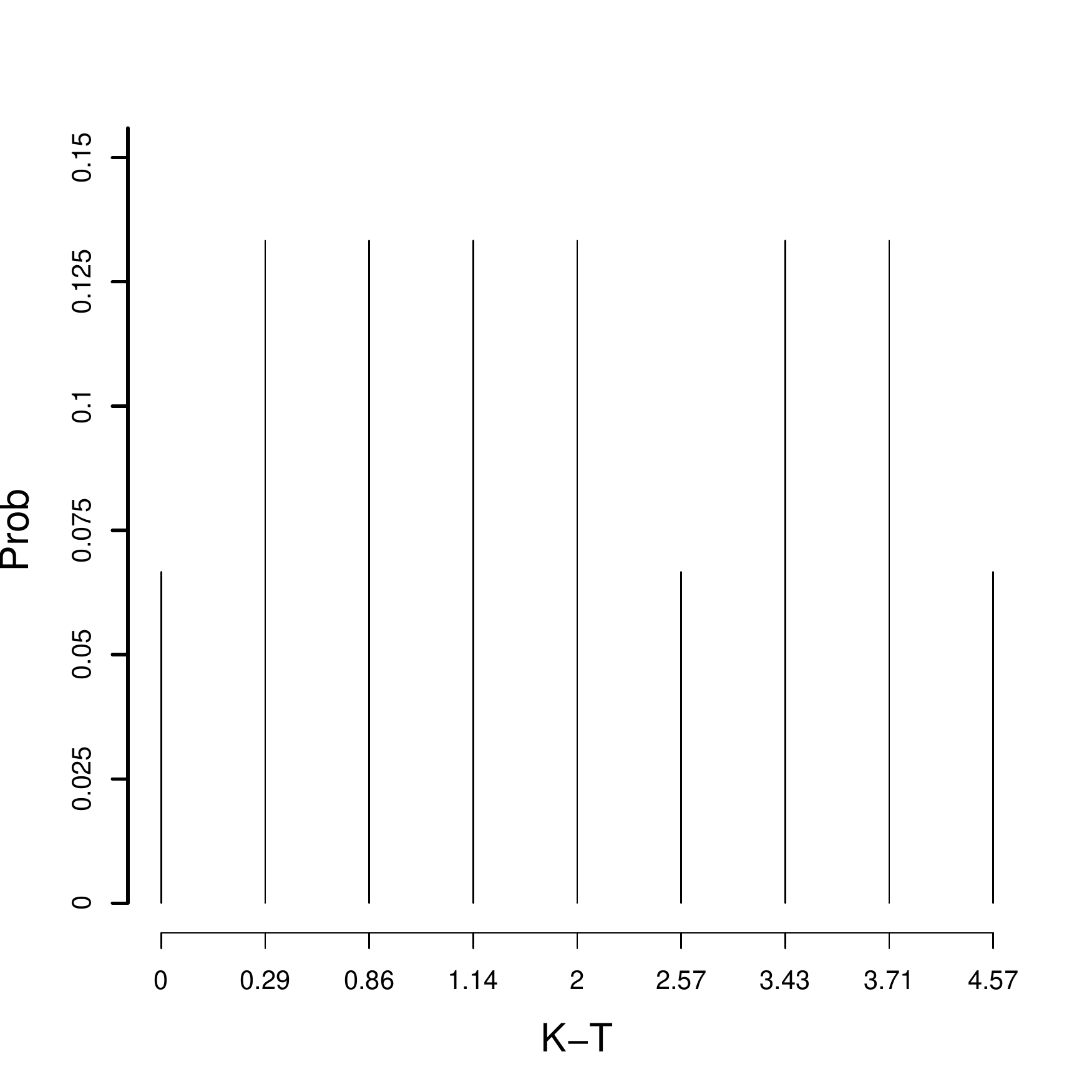} \\
\end{tabular}
\caption{Probability distribution for the Kruskal-Wallis  statistic  (KW) and  the -$\tau$ Concordance coefficient ($\tau$=1- relative dissorder)} 
\label{distprob_222}
\end{figure}

Table \ref{t:prob_DK_222} shows the probability distribution of the disorder and the  Concordance coefficient for 3 treatments with 2 patients each,  and Figure  \ref{distprob_222} compares the probability distribution of the Kruskal-Wallis statistic and the   Concordance coefficient,  for 3 treatments with 2 patients each.

\setlength{\tabcolsep}{7.3mm}
\renewcommand{\arraystretch}{.85}

\begin{table}[t] 	\centering\footnotesize
\begin{tabular}{ccc} 
    $dis$ & $\tau$   &$ Prob$           \\ \toprule
0	&	1.0000	&	0.06667	\\ \cmidrule(l){1-3}
1	&	0.8333	&	0.13333	\\ \cmidrule(l){1-3}
2	&	0.6667	&	0.20000	\\ \cmidrule(l){1-3}
3	&	0.5000	&	0.20000\\ \cmidrule(l){1-3}
4	&	0.3333	&	0.20000 	\\ \cmidrule(l){1-3}
5	&	0.1667	&	0.13333	\\ \cmidrule(l){1-3}
6	&	0.0000	&	0.06667\\ \bottomrule[0.2mm]
\end{tabular}
\caption{Probability distribution for the disorder ($dis$) and  the Concordance  coefficient $\tau$}
\label{t:prob_DK_222}
\end{table}

\section{Kendall Concordance Test }

In this section, we present the  Concordance Test  in order to evaluate when  different samples come from the same population   distribution.  The randomization test  introduced by \cite{Fisher1935} establishes a framework for the statistical test based on permutations, see also \citep{Box1980,Stern1990,Welch1990}.  

If all the samples provide from the same distribution then  all possible ways to rank $n$ observations  divided into $k$ samples have the same probability of occurring. If a result of the experiment provides an order of the observations with a high disorder, it will support the idea that all observations come from the same population, on the contrary, a result with a small disorder will go against the claim that the observations come from the same population. In this way, we propose to consider samples that come from the same distribution as null hypothesis, while the alternative hypothesis is that some of the samples come from another distribution.


\begin{description}
\item[$H_0$:] There is no difference among the $k$ populations.
\item[$H_a$:]  At least one of the populations differs from the other populations. 
\end{description}

The decision rule   is  to  reject the null hypothesis  if the disorder in the permutation of observations is small, equivalently if the concordance coefficient $\tau$ is close to one. We reject the null hypothesis $H_0$ at the significance level $\alpha$ if $\tau$ is greater than the percentile $(1-\alpha)\%$  of probability distribution  of $\tau$. 

The following example illustrates the use of the Concordance test proposed in this work and compares it with the classical Kruskal-Wallis non-parametric test . 

\setlength{\tabcolsep}{2.3mm}
\renewcommand{\arraystretch}{1.}
\begin{exmp}Suppose you have applied  three  treatments to 18 patients, measuring the the number of hours it takes these patients to recover.\\

\centering
\begin{tabular}{ccccccccccccc} \toprule
 & \multicolumn{10}{c}{Hours}\\ \hline
Treatment A & 12 & 13 & 15 & 20& 23 & 28 & 30 & 32 & 40 & 48 \\ \hline
Treatment B & 29& 31& 49 & 52 & 54 & & & & & \\ \hline
Treatment C & 24 & 26 & 44 & & & & & & &\\ \bottomrule[0.2mm]
\end{tabular}
\end{exmp}

\subsection*{Concordance Test}

The experiment  ranks the patients in the following ranking $$ (a\ a\ a\ a\ a\ c\ c\ a\ b\ a\ b\ a\ a\ c\ a\ b\ b\ b).$$

If we perform the contrast using the disorder statistic or the correlation coefficient $\tau$ , we must calculate the permutation of treatments that maximizes the order between patients obtained in the experiment. The matrix of preferences between treatments observed is as follows:


\begin{center}	\begin{tabular}{cccc}
		&\,\,\, \,\, $A$ & $B$ & $C$\,\,\,\, \\ 
		$A$&\multirow{3}{*}{  $ \left(\begin{array}{r}  $--$ \\7 \\  11 \end{array}\right.$}  &\multirow{3}{*}{  $ \begin{array}{c}43   \\ $--$  \\  13\end{array}$}   &\multirow{3}{*}{  $ \left.\begin{array}{c}19   \\2  \\  $--$ \end{array}\right)$}  \\
		$B$&&&\\
		$C$&&&\\
	\end{tabular}\end{center}

 The order between treatments that maximizes the order between patients corresponds to the order $(A\ C\ B)$,  satisfying 75 of the 95 preferences contained in the matrix, where the value 75 is the solution of the Linear Ordering Problem (LOP)\footnote{The solution of LOP for this example is the permutation of sets that maximizes the preferences in the preference  matrix. It is sufficient to compare the 6 possibilities, $(A\ B\ C) = 64 $, $(A\ C\ B) = 75 $, $(B\ A\ C) = 28$, $(B\ C\ A) = 20$,$(C\ A\ B) = 67$ and $(C\ B\ A) = 31$. }. 
 Therefore exactly 20 = 95-75 is the number of pairwise disagreements necessary to order the samples and obtain the order (ACB), that is the disorder is 20. The greatest disorder that an order of elements can have with samples of 10, 5 and 3 elements is given by:
  $\sum_{r<s} \lfloor \frac{n_r\, n_S}{2} \rfloor + GP_b=  47 + 1 = 48 $,  therefore the concordance coefficient is  $\tau = 1-20/48= 0.5744 $.   The p-value of the disorder 20 or equivalently  of the  concordance coefficient $\tau=0.5744$ is 0.0492725\footnote{Tables of p-values for the Concordance coefficient $\tau$ are in  \ref{A:3}},  therefore, at a level of significance less than 5\% we can reject the null hypothesis of equality in treatments. 


\subsection*{Kruskal-Wallis Test}

The treatments  A, B, and C have average ranks of 7.3, 14.2 and 9 respectively, and the sum of ranks $R_A=73$, $R_B=71$ and $R_C=27$ respectively.

The Kruskal-Wallis statistic is given by:

\[ K = -3(n+1)+\frac{12}{n(n+1)}\sum \frac{R_i^2}{n_i}= -3(18+1)+\frac{12}{19(19+1)}\left( \frac{73^2}{10} +\frac{71^2}{5}+\frac{27^2}{3}\right)=5.6\]

 In \citep{Meyer2015} the exact values for the Kruskall Wallis contrast and different levels of significance are found. We can conclude by looking at the tables that the p-value of the K statistic is greater than 0.05, therefore, we cannot reject the null hypothesis that the treatments are equally effective.

\begin{figure}[htb!]\centering
\begin{tabular}{cc}
		\includegraphics[width=5.5cm]{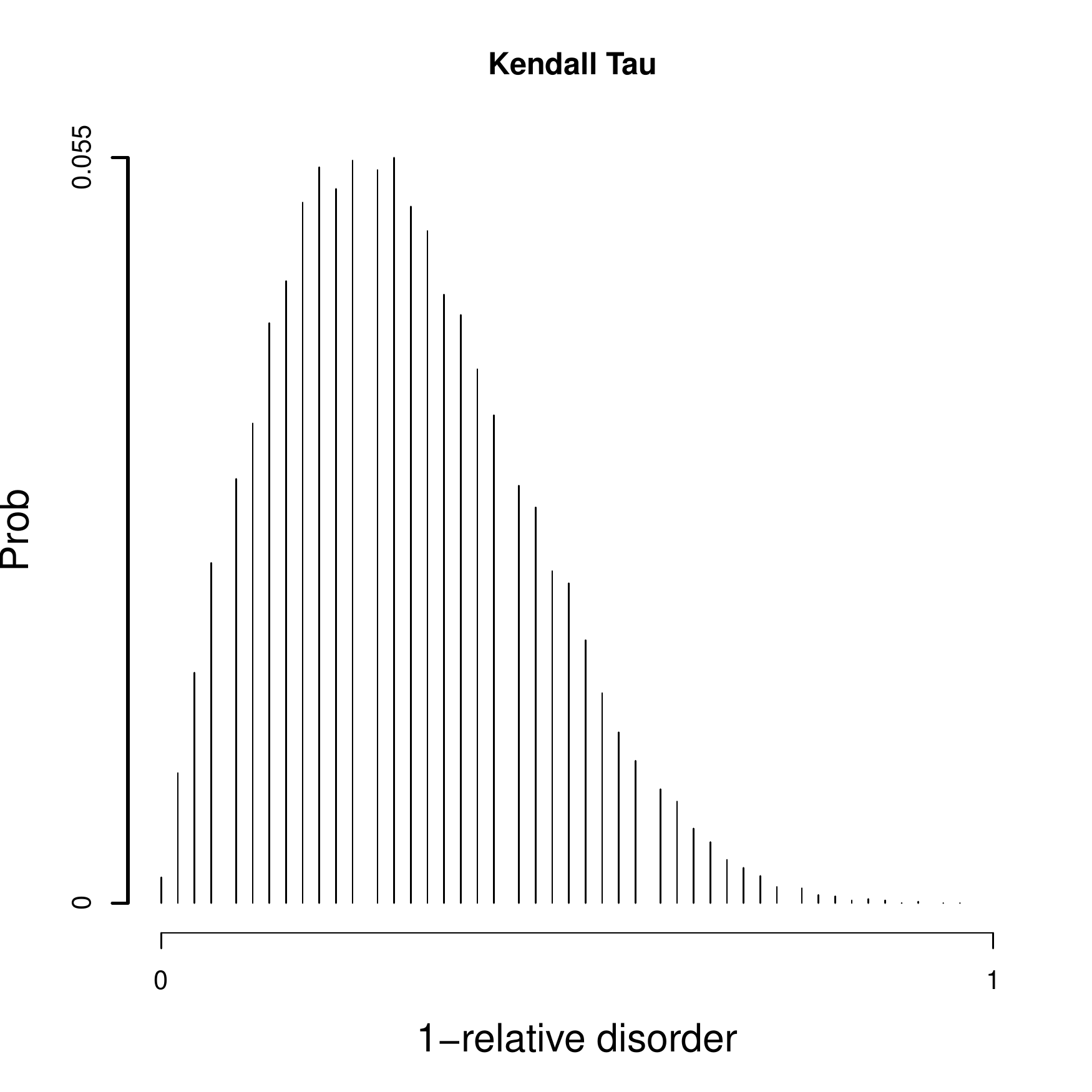}  &\includegraphics[width=5.5cm]{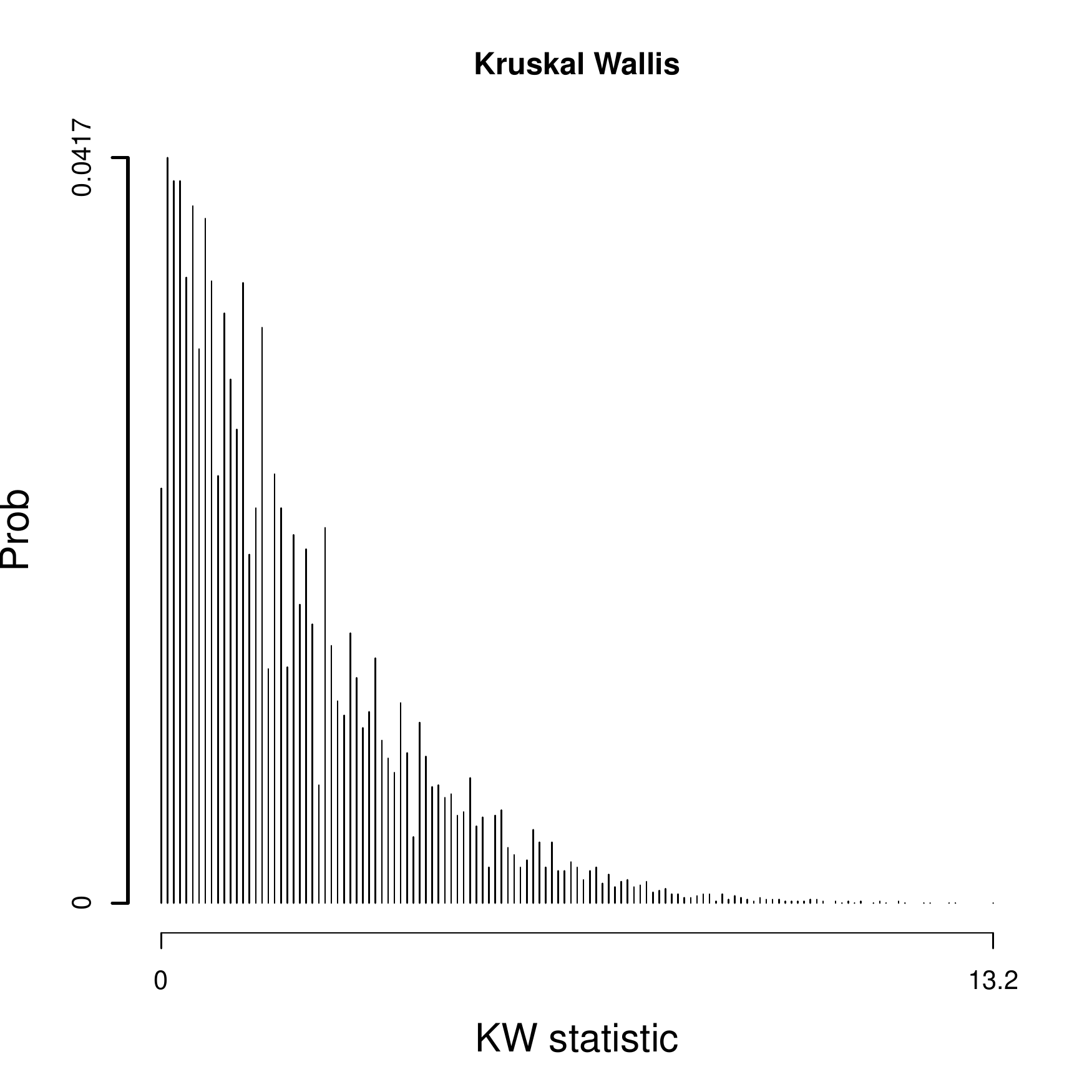} \\
\end{tabular}
\caption{Probability Distribution functions for the Concordance coefficient ($\tau$=1- disorder)  and   Kruskal-Wallis   ($KW$) statistics, with sample size  $N=(10,5,3)$} 
\label{f:prob_1053}
\end{figure}

The comparison of both methods,   Concordance and the  Kruskal-Wallis test providew similar results about the statistic but the conclusion  differs. Figure \ref{f:prob_1053} shows the exact probability function for both distributions, and the first conclusion is that  the Concordance statistic $\tau$ presents a more symmetric distribution than the  Kruskal-Wallis statistic.  \ref{A:2} presents the density probability functions for several experiments, where  sample sizes varying form $N=(4,4)$ to $N=(5,5,4,4,4,4,4)$.

%
%

\subsection*{Concordance  Test with ties}

\setlength{\tabcolsep}{2.3mm}
\renewcommand{\arraystretch}{1.}
\begin{exmp}[continuation] Suppose you have the same experiment but with ties. Ties are in bold.\\
\centering 

\centering
\begin{tabular}{ccccccccccccc} \toprule
 & \multicolumn{10}{c}{Hours}\\ \hline
Treatment A & 12 & 13 & 15 & 20& {\bf 24} & {\bf 29} & 30 & 32 & 40 & {\bf 49} \\ \hline
Treatment B & {\bf 29}& 31& {\bf 49} & 52 & 54 & & & & & \\ \hline
Treatment C & {\bf 24} & 26 & 44 & & & & & & &\\ \bottomrule[0.2mm]
\end{tabular}
\end{exmp}

The results of the experiment order the individuals according to the sequence: 

$$ (a\ a\ a\ a\ (a\ c)\ c\ (a\ b)\ a\ b\ a\ a\ c\ (a\ b)\ b\ b) $$

where the elements grouped in the order indicates that they tie.   There are  $8$ different possibilities  in order to undo  ties in the ranking of elements, if the same probability is assumed for all of them, the expected preference matrix between treatments is given distributing with the same weight the preference in the comparison of repeated observations, that is, assigning the value 0.5 to each of the treatments when two tied units are compared. The preference matrix for this example would be as follows:


\begin{center}	\begin{tabular}{cccc}
		&\,\,\, \,\, $A$ & $B$ & $C$\,\,\,\, \\ 
		$A$&\multirow{3}{*}{  $ \left(\begin{array}{r}  $--$ \\8 \\  11.5 \end{array}\right.$}  &\multirow{3}{*}{  $ \begin{array}{c}42   \\ $--$  \\  13\end{array}$}   &\multirow{3}{*}{  $ \left.\begin{array}{c}18.5   \\2  \\  $--$ \end{array}\right)$}  \\
		$B$&&&\\
		$C$&&&\\
	\end{tabular}\end{center}
	
Note that the previous matrix represents the matrix of expected preferences if all permutations of items with ties in which they are undone are considered, with the same probability of tie between elements.


The order between treatments that maximizes the order between patients corresponds to the order $(A\ C\ B)$, satisfying 73.5 of the 95 preferences contained in the matrix, where 73.5 is the solution of the linear ordering problem. Therefore, 21.5 = 95-73.5 is the expected number of pairwise disagreements necessary to order the samples and obtain the order $(A\ C\ B)$,  that is the disorder is 21.5 or equivalent and the concordance coefficient is $\tau = 1-21.5/48= 0.5521 $, a value with a significance greater than 0.05,
$p - value > 0.05$.  In this case the observed data do not show significant evidence 
in favor of a difference in the effectiveness of treatments.


\subsection*{Kruskal-Wallis Test with ties}

The treatments  A, B, and C have average ranks of 7.45, 14 and 8.83 respectively, and the sum of ranks $R_A=74.5$, $R_B=70$ and $R_C=26.5$ respectively.

The Kruskal-Wallis statistic is given by:

\[ K = -3(n+1)+\frac{12}{n(n+1)}\sum \frac{R_i^2}{n_i}= -3(18+1)+\frac{12}{19(19+1)}\left( \frac{74.5^2}{10} +\frac{70^2}{5}+\frac{26.5^2}{3}\right)=5.074\]

If we make the adjustment in the statistic for ties, we get:

\[ \tilde{K} = \frac{K}{\displaystyle 1-\frac{ \sum_{i=1}^3(t_i^2-t_i)}{N^3-N}} = \frac{5.074}{\displaystyle 1-\frac{(2^2-2)+(2^2-2)+(2^2-2)}{18^3-18}}=\displaystyle 5.391\]
 
In this case, the Kruskal-Wallis provides the same conclusion as the Concordance test; uncertainty being greater when we have ties.

\section{Computing p-values}

In order to compute the  probability distribution for the Concordance coefficient  statistic, the enumeration of all the permutations of elements from a order is required. Note for example that if  we  have 4 samples with 6 elements each $N=(6,6,6,6)$, the number of possible results in the experiment is $24!/6!6!6!6!$ $= 2.15433\cdot 10 ^{20}$. The total computational time to compute the Concordance coefficient for all $2.15433\cdot 10 ^{20}$ possibilities   was  more than 60 days in n Intel  Intel(R) Xeon (R)   processor CPU E5-2650 v3 @ 2.30 GHz, 20 cores and RAM 64 GiB.  Algorithm 1 presents the recursive function used to evaluate the concordance coefficient probability distribution. 

 \ref{A:3} presents the critical values and  exact p-values for three different significance levels, 0.10, 0.05 and 0.01 and different sample sizes.

\begin{algorithm}  \small
\setstretch{1.1}
  \KwData{\\
  $\mathbf{p}:$ ordered array of integers  with ties,\\
  $n:$ length of $\mathbf{p}$,\\
  $s=0:$ start to the   the permutation.}
\DontPrintSemicolon
  \SetKwFunction{FMain}{}
  \SetKwProg{Fn}{Main}{}{}
  \Fn{\FMain{$\mathbf{p}$}}{
        Permutation($\mathbf{p}$,0,$n$); \;
         \KwRet\;
  }
  \SetKwFunction{FPerm}{}
  \SetKwProg{Fn}{Permutation}{:}{}
  \Fn{\FPerm{$\mathbf{p}$,$s$,$n$}}{
         Dissorder($\mathbf{p}$); \;
    	int $tmp=0$;\;
	\If{$s<n$}{
	\For{$i=n-2: \ i\geq s ;\, i--$}{
	 \For{$j=i+1;\, j<n;\, j ++$}{
	\If{$p[i]\neq p[j]$}{
	   $tmp = p[i]$;
            $p[i] = p[j]$;
            $p[j] = tmp$
            Permutation($\mathbf{p}$,$i+1$,$n$); \;
	}
	 }
	tmp=ps[i];\;
	 \For{$j=i+1;\, j<n;\, j ++$}{
	 $p[k]=p[k++]$;\;
	 }
	 $p[n-1]=tmp$;\;
	}
}
        \KwRet\;
  }
  \SetKwFunction{FDiss}{}
  \SetKwProg{Fn}{Dissorder}{}{}
   \Fn{\FDiss{$\mathbf{p}$}}{
\tcc{Evaluate the disorder and the Concordance coefficient  of  permutation $\mathbf{p}$}
        \KwRet\;
  }
  	\caption{Program  to compute  exact p-values of the Concordance coefficient statistic $\tau$. }
\end{algorithm}

\section{Conclusions and future research plans}

A new measure to estimate the Concordance coefficient of different samples is presented in this work, and a statistical test to determine when different observations come from the same distribution.  A comparison with a classical Kruskal-Wallis test was introduced   to show that both tests differ.

This work  aims to be   an introduction of the new concordance measure  between samples,  but there still remains much to be done. There is a  a new problem and further challenges for researchers , for example: study the asymptotic   distribution of the Concordance coefficient, exploring  the possibility of finding the exact distribution with the help of modern computing, analyzing  the power of the Concordance test presented in this work, and  finally, implementing the concordance coefficient  in a statistical  software such as R.

\section*{Acknowledgments}

The authors are grateful for the financial support from the Spanish Ministry for Economy and Competitiveness (Ministerio de Econom\'\i{}a, Industria y Competitividad), the State Research Agency (Agencia Estatal de Investigaci\'on) and the European Regional Development Fund (Fondo Europeo de Desarrollo Regional) under grant MTM2016-79765-P (AEI/FEDER, UE).


\newpage

\begin{appendix}

\section{Concordance coefficient and Kruskal-Wallis statistic for all possible results in an experiment with three treatments and  two people in each treatment} \label{A:1}

\setlength{\tabcolsep}{.5mm} 
\renewcommand{\arraystretch}{.001}
\renewcommand{\arraystretch}{.85}

\begin{table}[h] 	\centering \footnotesize
\begin{tabular}{cccccccccccccccccccccccccccccccccccccccccc}  
 &&&&&&&& $dis$ &&  $\tau$ &$KW$&&&&&&&&&&&& $diss$ &&  $\tau$ & $KW$&&&&&&&&&&&&  $dis$ &&  $\tau$ &$KW$\\
\toprule
	&	a	&	a	&	b	&	b	&	c	&	c	&&	0	&&	1.0000	&	4.57	&&	&&	&	b	&	a	&	a	&	b	&	c	&	c	&&	2	&&	0.6667	&	3.43	&&	&&	&	c	&	a	&	a	&	b	&	b	&	c	&&	4	&&	0.3333	&	1.14	\\	\cmidrule(l){1-12}		\cmidrule(l){16-27}		\cmidrule(l){31-42}
	&	a	&	a	&	b	&	c	&	b	&	c	&&	1	&&	0.8333	&	3.71	&&	&&	&	b	&	a	&	a	&	c	&	b	&	c	&&	3	&&	0.5000	&	2.00	&&	&&	&	c	&	a	&	a	&	b	&	c	&	b	&&	3	&&	0.5000	&	2.00	\\	\cmidrule(l){1-12}		\cmidrule(l){16-27}		\cmidrule(l){31-42}
	&	a	&	a	&	b	&	c	&	c	&	b	&&	2	&&	0.6667	&	3.43	&&	&&	&	b	&	a	&	a	&	c	&	c	&	b	&&	4	&&	0.3333	&	1.14	&&	&&	&	c	&	a	&	a	&	c	&	b	&	b	&&	2	&&	0.6667	&	3.43	\\	\cmidrule(l){1-12}		\cmidrule(l){16-27}		\cmidrule(l){31-42}
	&	a	&	a	&	c	&	b	&	b	&	c	&&	2	&&	0.6667	&	3.43	&&	&&	&	b	&	a	&	b	&	a	&	c	&	c	&&	1	&&	0.8333	&	3.71	&&	&&	&	c	&	a	&	b	&	a	&	b	&	c	&&	5	&&	0.1667	&	0.29	\\	\cmidrule(l){1-12}		\cmidrule(l){16-27}		\cmidrule(l){31-42}
	&	a	&	a	&	c	&	b	&	c	&	b	&&	1	&&	0.8333	&	3.71	&&	&&	&	b	&	a	&	b	&	c	&	a	&	c	&&	2	&&	0.6667	&	2.57	&&	&&	&	c	&	a	&	b	&	a	&	c	&	b	&&	4	&&	0.3333	&	0.86	\\	\cmidrule(l){1-12}		\cmidrule(l){16-27}		\cmidrule(l){31-42}
	&	a	&	a	&	c	&	c	&	b	&	b	&&	0	&&	1.0000	&	4.57	&&	&&	&	b	&	a	&	b	&	c	&	c	&	a	&&	3	&&	0.5000	&	2.00	&&	&&	&	c	&	a	&	b	&	b	&	a	&	c	&&	6	&&	0.0000	&	0.00	\\	\cmidrule(l){1-12}		\cmidrule(l){16-27}		\cmidrule(l){31-42}
	&	a	&	b	&	a	&	b	&	c	&	c	&&	1	&&	0.8333	&	3.71	&&	&&	&	b	&	a	&	c	&	a	&	b	&	c	&&	4	&&	0.3333	&	0.86	&&	&&	&	c	&	a	&	b	&	b	&	c	&	a	&&	5	&&	0.1667	&	0.29	\\
\end{tabular}
	\caption{All possible results in the experiment with samples sizes N=(2,2,2) }
       \label{samplespace222}
\end{table}

\setlength{\tabcolsep}{1.mm}
\renewcommand{\arraystretch}{.95}

\clearpage
\section{Comparasion of Distributions} \label{A:2}

\scriptsize
\begin{longtable}{|ccc|} \hline 
\caption{Probability density  function   of  the Kruskal-Wallis (dashed lines)  and the Concordance coefficient (continuous lines) statistics generated by simulation. Number of simulations 100,000. Note that the KW statistic has been normalized between 0 and 1. }\\
\endfirsthead

\multicolumn{3}{c}%
{{\bfseries \tablename\ \thetable{} -- continued from previous page}}  \\ \cline{1-3}
\endhead

\hline \multicolumn{3}{r}{{Continued on next page}} \\ \hline
\endfoot

\hline 
\endlastfoot

\hline
		\includegraphics[width=4.05cm]{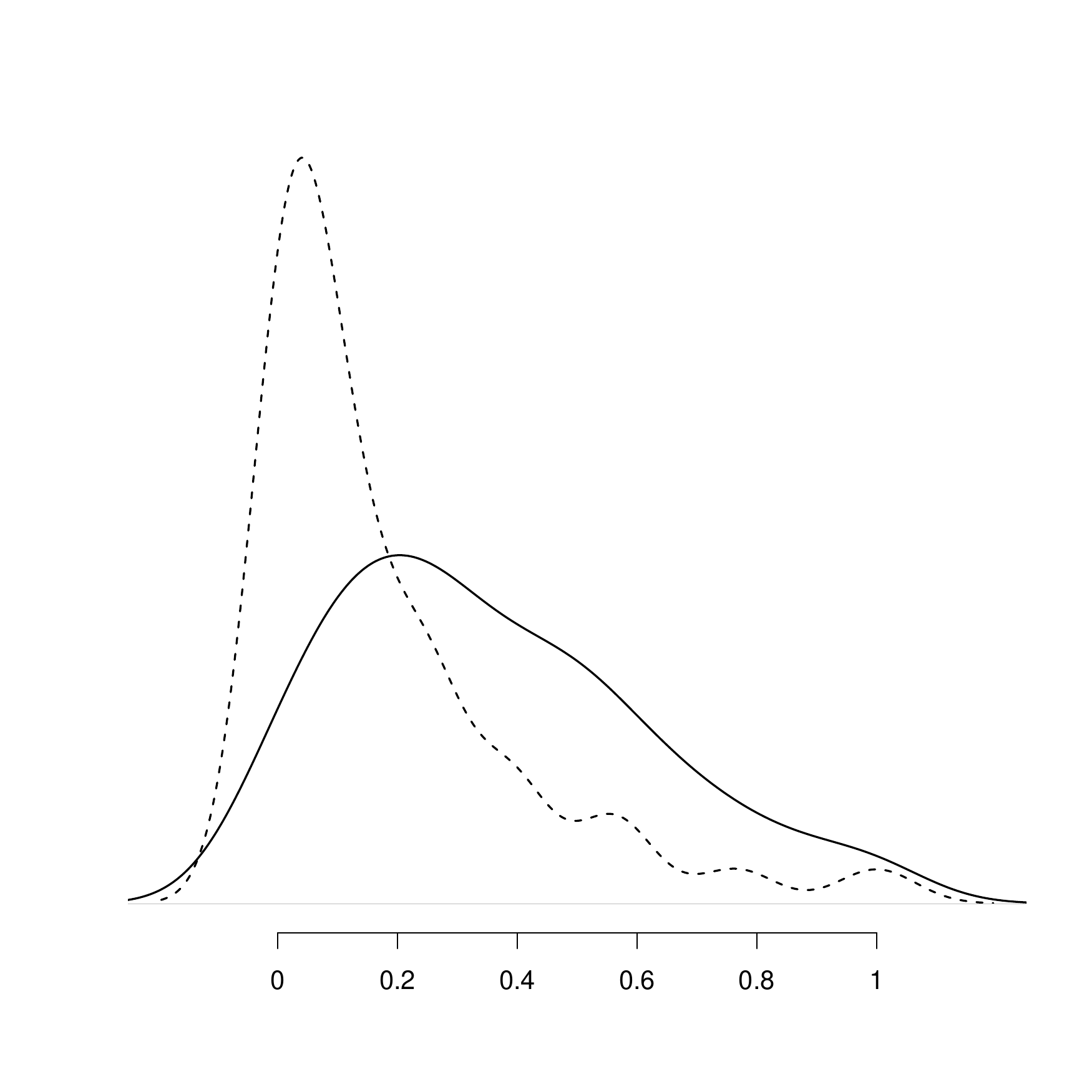} & 
		\includegraphics[width=4.05cm]{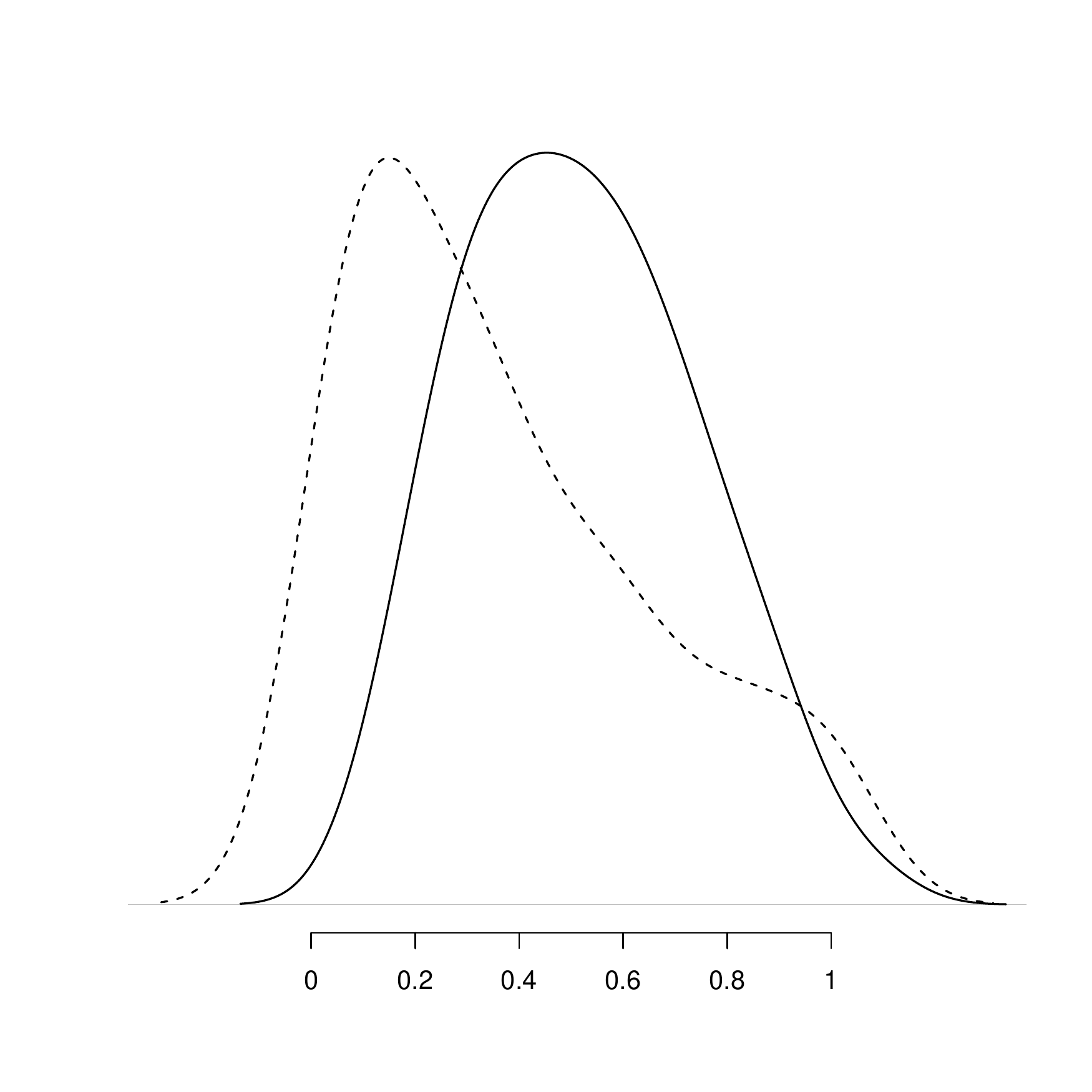} & 
		\includegraphics[width=4.05cm]{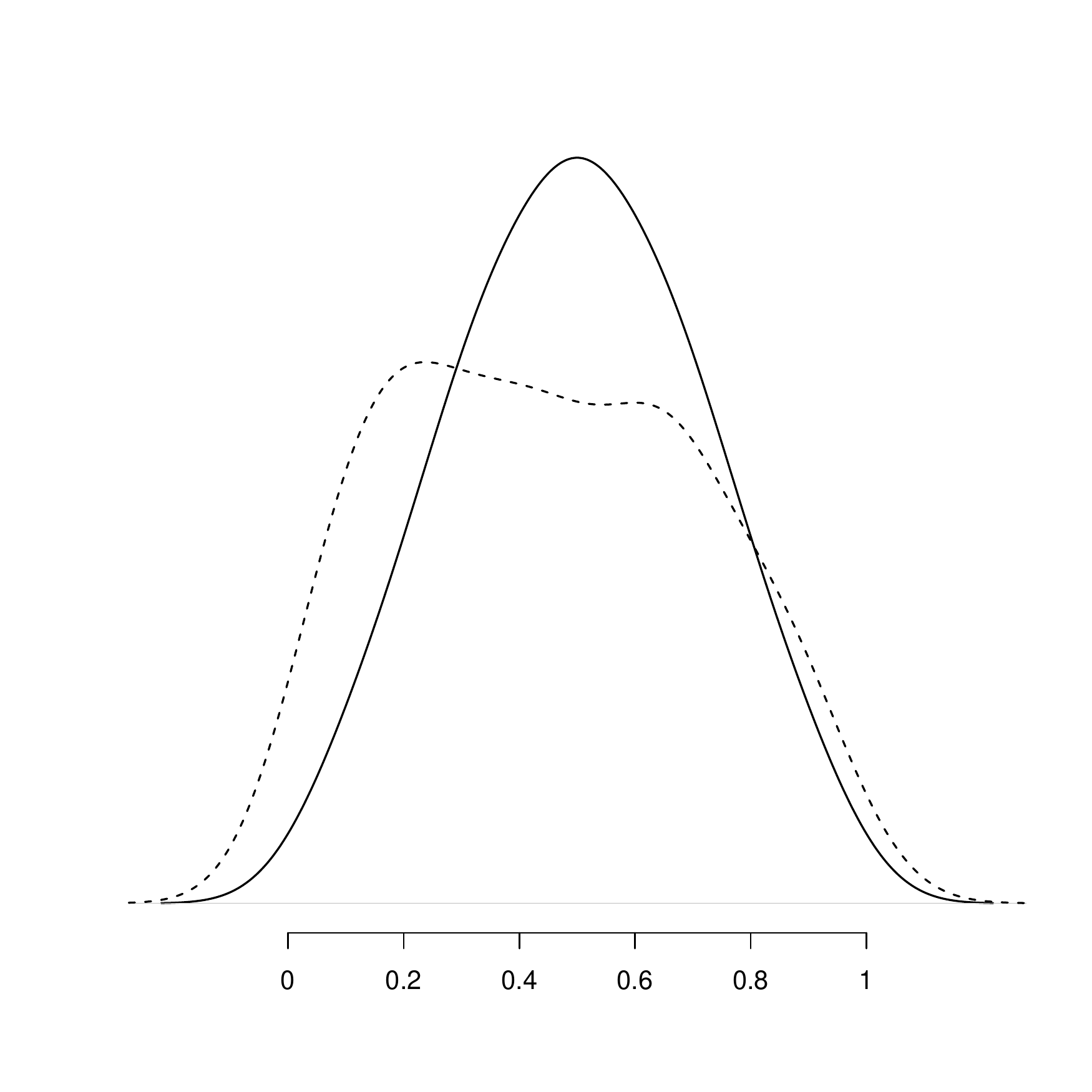} \\ 
		           Sample\_Size=(4,4)  &  Sample\_Size=(3,3,2)  &   Sample\_Size=(2,2,2,2) \\  
\hline 
	\includegraphics[width=4.05cm]{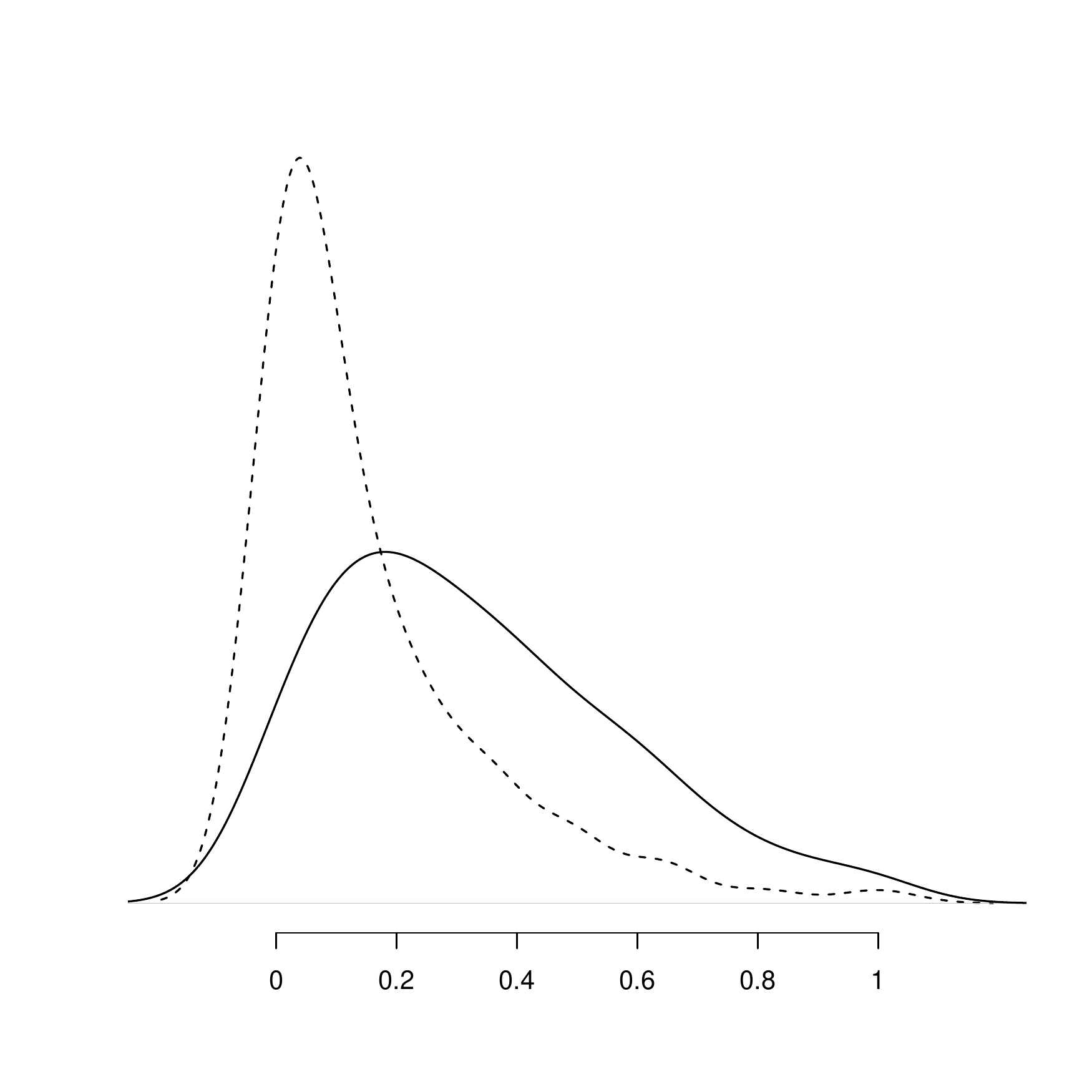} & 
		\includegraphics[width=4.05cm]{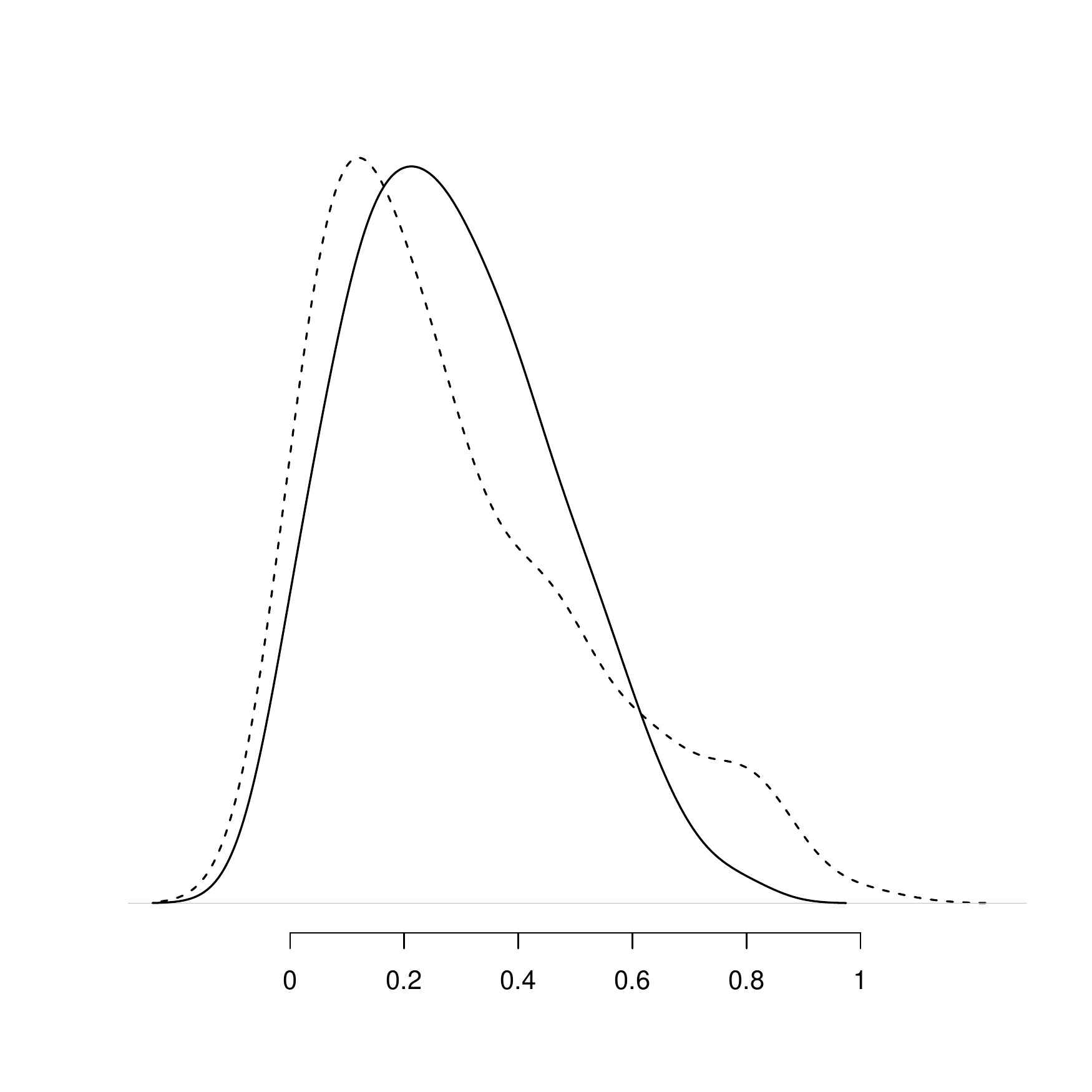} & 
		\includegraphics[width=4.05cm]{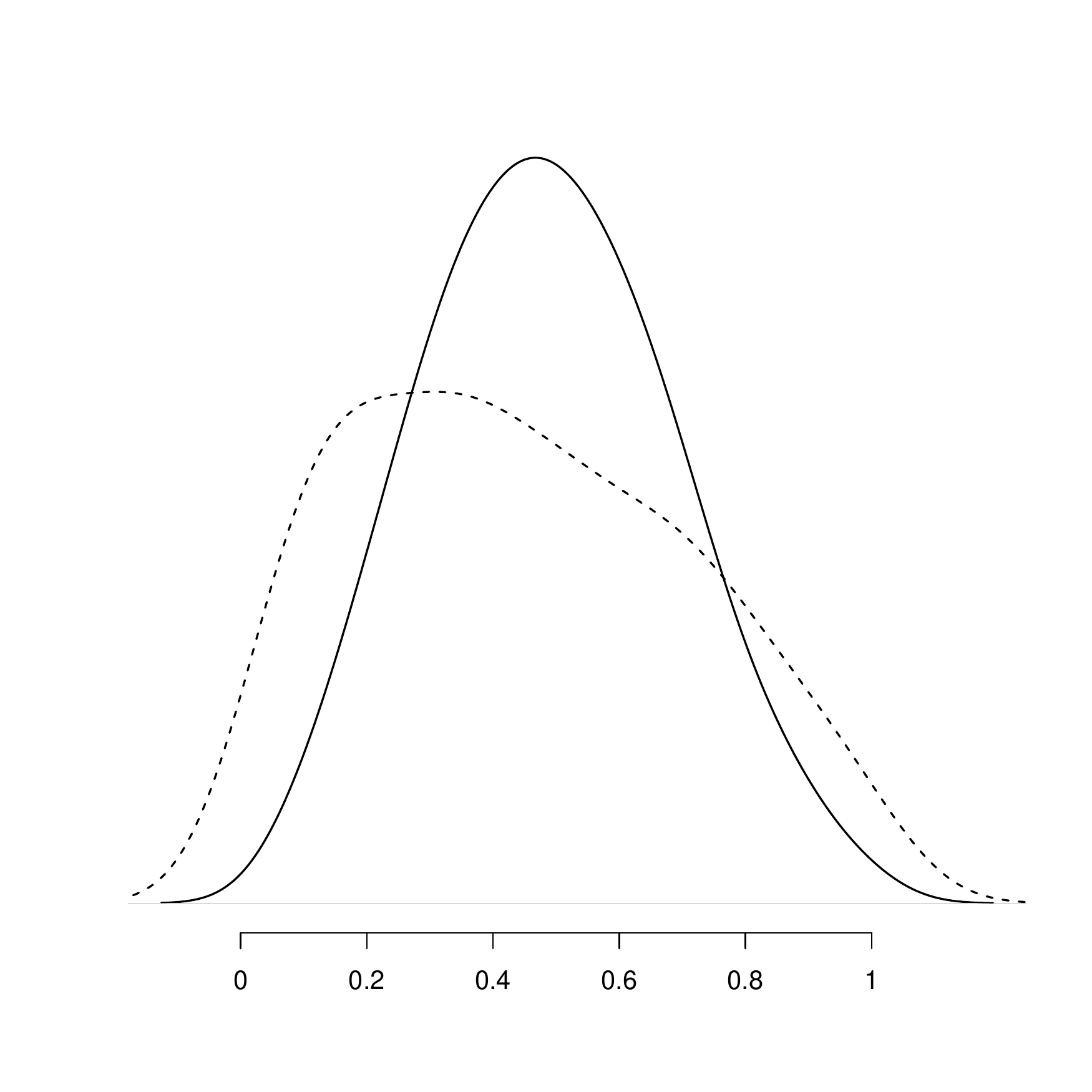} \\ 
		              Sample\_Size=(5,4) &   Sample\_Size=(3,3,3)  &    Sample\_Size=(3,2,2,2) \\  
\hline 
	\includegraphics[width=4.05cm]{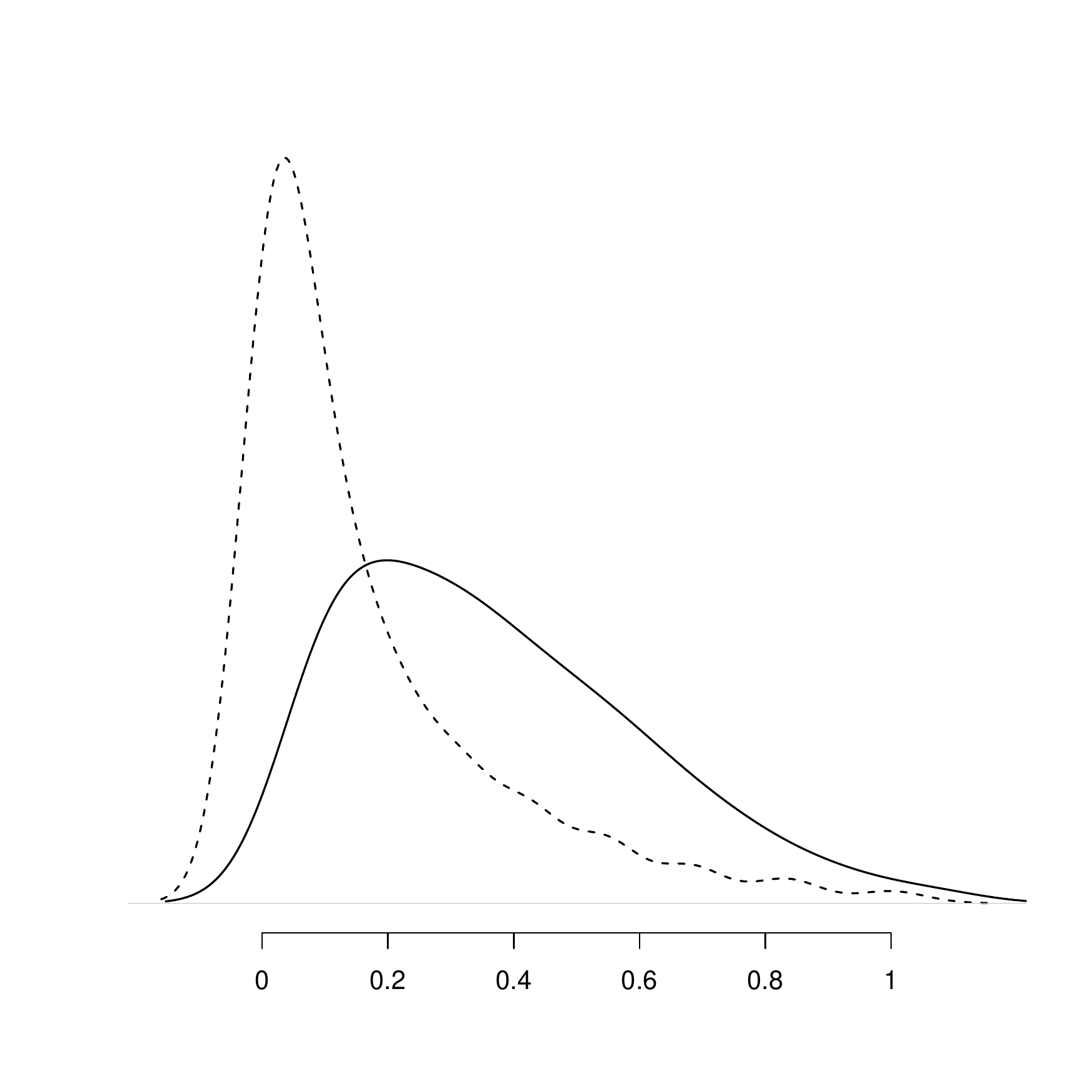} & 
		\includegraphics[width=4.05cm]{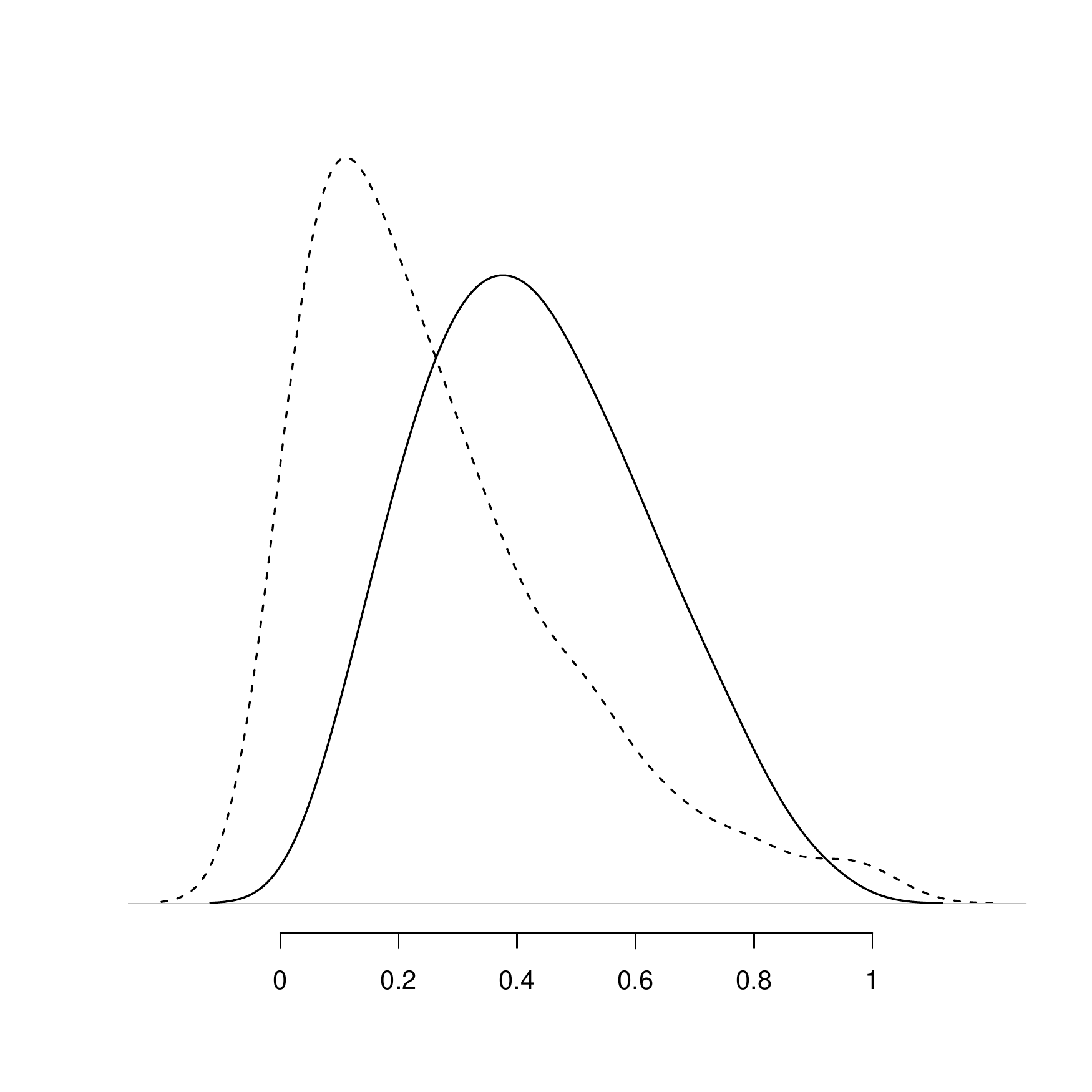} & 
		\includegraphics[width=4.05cm]{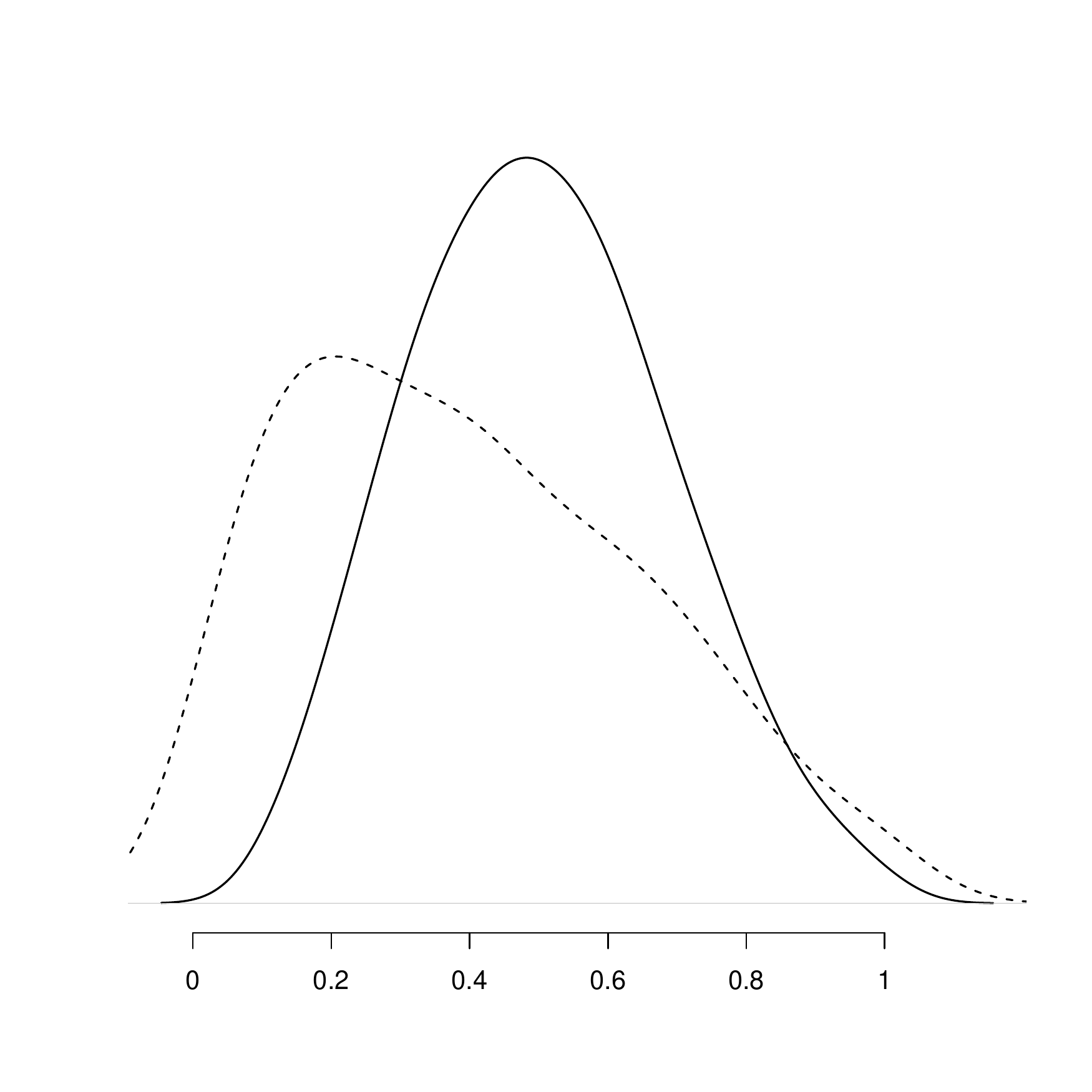} \\ 
		              Sample\_Size=(5,5) &   Sample\_Size=(4,3,3)  &    Sample\_Size=(3,3,2,2) \\  
		              \hline

		\includegraphics[width=4.05cm]{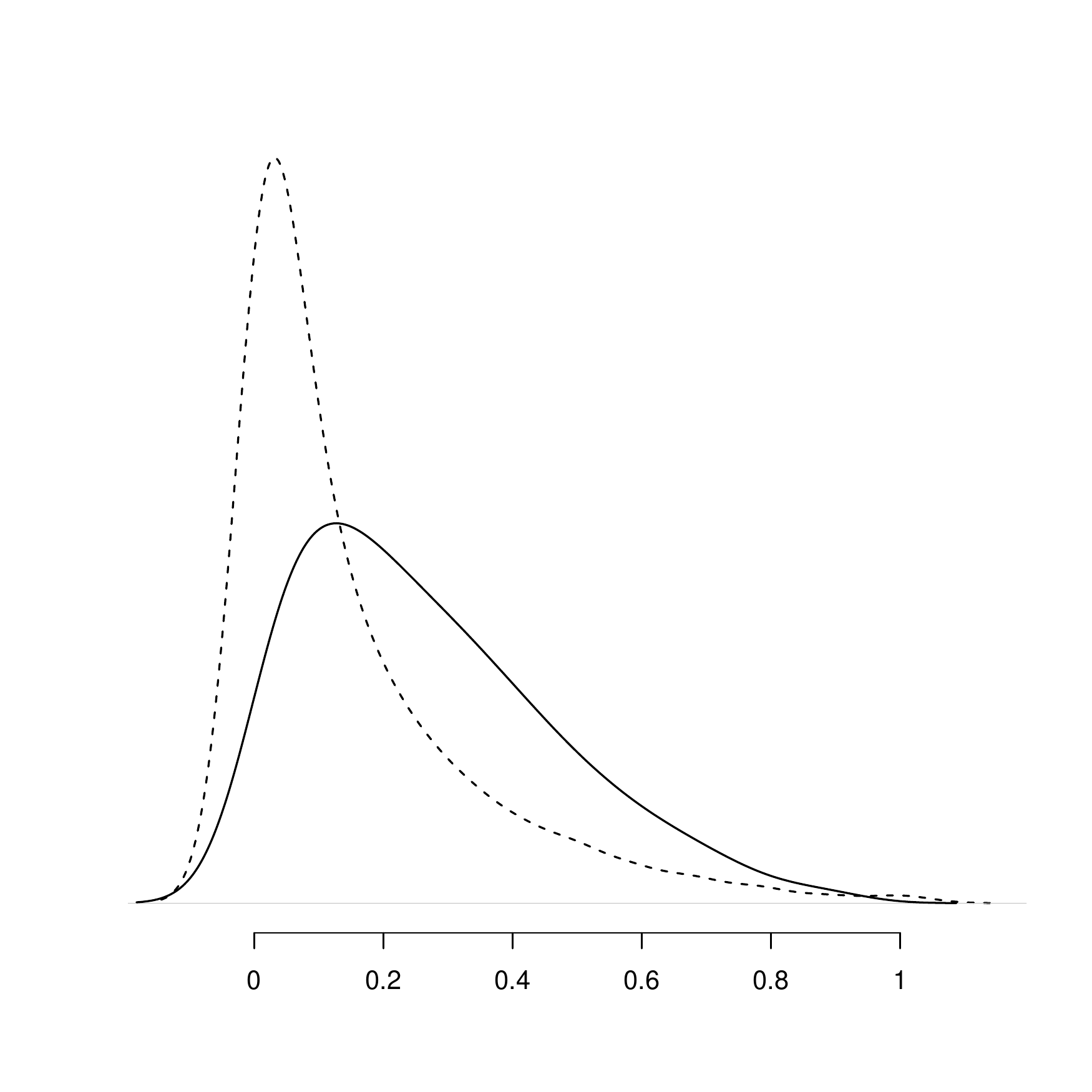} & 
		\includegraphics[width=4.05cm]{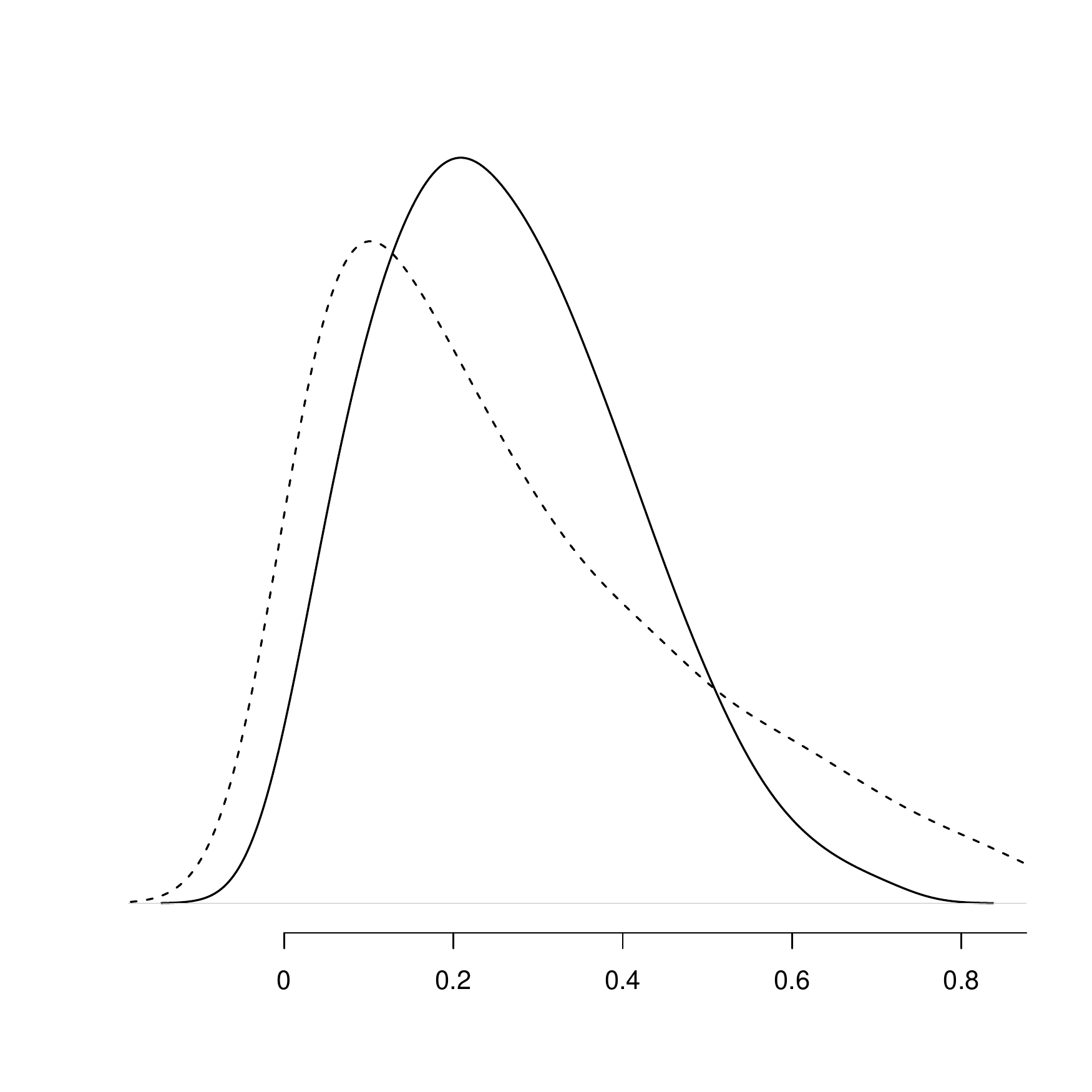} & 
		\includegraphics[width=4.05cm]{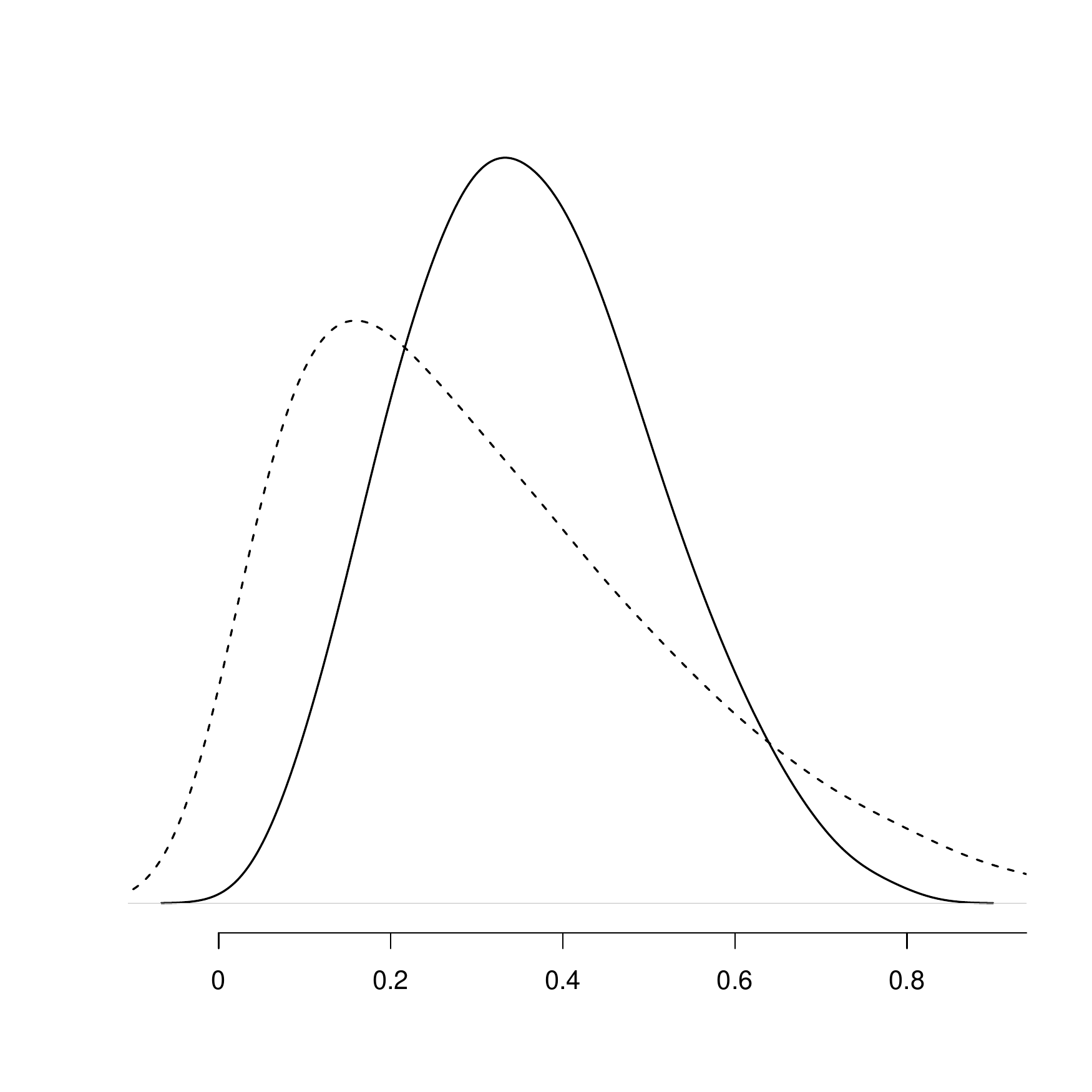} \\ 
		              Sample\_Size=(7,6) &   Sample\_Size=(5,5,5)  &    Sample\_Size=(4,4,4,3) \\
\end{longtable}

\newpage
\section{Kendall Tau Tables} \label{A:3}

\setlength{\tabcolsep}{1.mm}
\renewcommand{\arraystretch}{.85}

\begin{longtable}[h]{rrrcccccccccccc}
\caption{Critical values and Exact p-values at Desired Significance Levels of 0.10. 0.05 and 0.01 for k=2 samples. Concordance coefficient statistic $\tau$} \\ 
\endfirsthead

\multicolumn{15}{c}%
{{\bfseries \tablename\ \thetable{} -- continued from previous page}}  \\ \cline{1-15}
\hline \multicolumn{4}{c}{Samples Sizes} &  dis.  & $\tau$ \, & p-value &&  dis.  & $\tau$ \, & p-value &&  dis.  & $\tau$ \,& p-value \\ \hline
\endhead

\hline \multicolumn{15}{r}{{Continued on next page}} \\ \hline
\endfoot

\hline 
\endlastfoot
\hline
   & &&  &  \multicolumn{3}{c}{.10} &&  \multicolumn{3}{c}{.05} &&  \multicolumn{3}{c}{.01} \\  \cline{5-7} \cline{9-11} \cline{13-15}
\multicolumn{4}{c}{Sample Sizes} &  dis.  & $\tau$ \, & p-value &&  dis.  & $\tau$ \, & p-value &&  dis.  & $\tau$ \,& p-value \\ \hline

4	&	1	&&&		&		&		&&		&		&		&&		&		&		\\
4	&	2	&&&		&		&		&&		&		&		&&		&		&		\\
4	&	3	&&&	0	&	1.000000	&	0.057143	&&		&		&		&&		&		&		\\
4	&	4	&&&	1	&	0.875000	&	0.057143	&&	0	&	1.000000	&	0.028571	&&		&		&		\\
5	&	1	&&&		&		&		&&		&		&		&&		&		&		\\
5	&	2	&&&	0	&	1.000000	&	0.095238	&&		&		&		&&		&		&		\\
\hline
\end{longtable}

\newpage

\begin{longtable}[h]{rrrccccccccccccc}
\caption{Critical values and Exact p-values at Desired Significance Levels of 0.10. 0.05 and 0.01 for k=3 samples. Concordance coefficient statistic $\tau$} \\ 
\endfirsthead

\multicolumn{16}{c}%
{{\bfseries \tablename\ \thetable{} -- continued from previous page}}  \\ \cline{1-16}
\hline \multicolumn{5}{c}{Samples Sizes} &  dis.  & $\tau$ \, & p-value &&  dis.  & $\tau$ \, & p-value &&  dis.  & $\tau$ \,& p-value \\ \hline
\endhead

\hline \multicolumn{16}{r}{{Continued on next page}} \\ \hline
\endfoot

\hline 
\endlastfoot
\hline

   & && & &  \multicolumn{3}{c}{.10} &&  \multicolumn{3}{c}{.05} &&  \multicolumn{3}{c}{.01} \\  \cline{6-8} \cline{10-12} \cline{14-16}
\multicolumn{5}{c}{Sample Sizes} &  dis.  & $\tau$ \, & p-value &&  dis.  & $\tau$ \, & p-value &&  dis.  & $\tau$ \,& p-value \\ \hline
2	&	1	&	1	&&&		&		&		&&		&		&		&&		&		&		\\
2	&	2	&	1	&&&		&		&		&&		&		&		&&		&		&		\\
2	&	2	&	2	&&&	0	&	1.000000	&	0.066667	&&		&		&		&&		&		&		\\
3	&	1	&	1	&&&		&		&		&&		&		&		&&		&		&		\\
3	&	2	&	1	&&&		&		&		&&		&		&		&&		&		&		\\
3	&	2	&	2	&&&	1	&	0.875000	&	0.085714	&&	0	&	1.000000	&	0.028571	&&		&		&		\\
\hline
\end{longtable}

\newpage

\scriptsize
\begin{longtable}[h]{rrrcccccccccccccc}
\caption{Critical values and Exact p-values at Desired Significance Levels of 0.10. 0.05 and 0.01 for k=4 samples. Concordance coefficient statistic $\tau$} \\ 
\endfirsthead

\multicolumn{17}{c}%
{{\bfseries \tablename\ \thetable{} -- continued from previous page}}  \\ \cline{1-17}
\hline \multicolumn{6}{c}{Samples Sizes} &  dis.  & $\tau$ \, & p-value &&  dis.  & $\tau$ \, & p-value &&  dis.  & $\tau$ \,& p-value \\ \hline
\endhead

\hline \multicolumn{17}{r}{{Continued on next page}} \\ \hline
\endfoot

\hline 
\endlastfoot
\hline

   & && && &  \multicolumn{3}{c}{.10} &&  \multicolumn{3}{c}{.05} &&  \multicolumn{3}{c}{.01} \\  \cline{7-9} \cline{11-13} \cline{15-17}
\multicolumn{6}{c}{Sample Sizes} &  dis.  & $\tau$ \, & p-value &&  dis.  & $\tau$ \, & p-value &&  dis.  & $\tau$ \,& p-value \\ \hline
2	&	2	&	1	&	1	&&&		&		&		&&		&		&		&&		&		&		\\
2	&	2	&	2	&	1	&&&	0	&	1.000000	&	0.038095	&&	0	&	1.000000	&	0.038095	&&		&		&		\\
2	&	2	&	2	&	2	&&&	2	&	0.833333	&	0.095238	&&	1	&	0.916667	&	0.038095	&&	0	&	1.000000	&	0.009524	\\
3	&	1	&	1	&	1	&&&		&		&		&&		&		&		&&		&		&		\\
3	&	2	&	1	&	1	&&&	0	&	1.000000	&	0.057143	&&		&		&		&&		&		&		\\
3	&	2	&	2	&	1	&&&	1	&	0.909091	&	0.050000	&&	0	&	1.000000	&	0.014286	&&		&		&		\\
\hline
\end{longtable}

\end{appendix}

\end{document}